\documentclass[12pt]{article}

\usepackage{amsmath,amsfonts,amssymb}

\def\bea{\begin{eqnarray}}
\def\eea{\end{eqnarray}}
\def\be{\begin{equation}}
\def\ee{\end{equation}}
\def\ba{\begin{array}}
\def\ea{\end{array}}
\def\nn{\nonumber}

\font\tenrsfs=rsfs10
\font\sevenrsfs=rsfs7
\font\fiversfs=rsfs5
\newfam\rsfsfam
\textfont\rsfsfam=\tenrsfs
\scriptfont\rsfsfam=\sevenrsfs
\scriptscriptfont\rsfsfam=\fiversfs
\def\mathscr#1{{\fam\rsfsfam\relax#1}}
\def\a{& \hspace{-11pt}}

\begin{document}

\begin{titlepage} 

\rightline{\footnotesize{CERN-PH-TH/2009-040}} \vspace{-0.2cm}

\begin{center}

\vskip 0.4 cm

\begin{center}
{\Large{ \bf Globally and locally supersymmetric \\ [2mm] effective theories for light fields}}
\end{center}

\vskip 1cm

{\large Leonardo Brizi$^{a}$, Marta G\'omez-Reino$^{b}$ and Claudio A. Scrucca$^{a}$
}

\vskip 0.8cm

{\it
$^{a}$Institut de Th\'eorie des Ph\'enom\`enes Physiques, EPFL, \\
\mbox{CH-1015 Lausanne, Switzerland}\\
$^{b}$Theory Division, Physics Department, CERN, \\
CH-1211 Geneva 23, Switzerland\\
}

\vskip 0.8cm

\end{center}

\begin{abstract}

We reconsider the general question of how to characterize most efficiently the 
low-energy effective theory obtained by integrating out heavy modes in globally 
and locally supersymmetric theories. We consider theories with chiral and vector 
multiplets and identify the conditions under which an 
approximately supersymmetric low-energy effective theory can exist. These 
conditions translate into the requirements that all the derivatives, fermions and 
auxiliary fields should be small in units of the heavy mass scale. They apply 
not only to the matter sector, but also to the gravitational one if present, 
and imply in that case that the gravitino mass should be small. We then show how 
to determine the unique exactly supersymmetric theory that approximates 
this effective theory at the lowest order in the counting of derivatives, fermions 
and auxiliary fields, by working both at the superfield level and with component 
fields. As a result we give a simple prescription for integrating out heavy superfields 
in an algebraic and manifestly supersymmetric way, which turns out to hold in the 
same form both for globally and locally supersymmetric theories, meaning that 
the process of integrating out heavy modes commutes with the process of switching 
on gravity. More precisely, for heavy chiral and vector multiplets one has to impose 
respectively stationarity of the superpotential and the K\"ahler potential.

\end{abstract}

\bigskip

\end{titlepage}

\newpage

\section{Introduction} \setcounter{equation}{0}

A general problem commonly encountered in many different physical contexts 
is that of integrating out heavy fields in a supersymmetric theory, in order to 
define a low-energy effective theory for the remaining light fields. 
In general, the heavy fields will be stabilized at values implying a spontaneous 
breakdown of supersymmetry, and the low-energy effective theory will consequently 
be non-supersymmetric. In such a case, the best thing that one can do is to proceed 
in the same way as for ordinary effective theories. In particular, at the two-derivative 
level the effective theory is obtained by determining the heavy fields in terms of the 
light ones by requiring stationarity of the potential with respect to the heavy fields. 
However, it may happen that the heavy fields are stabilized in an approximately 
supersymmetric way, with vacuum expectation values that break only very little or 
not at all supersymmetry. One may then expect that the low-energy effective theory 
for the light fields should be approximately supersymmetric, and try to exploit this fact 
to describe it more efficiently. More precisely, at the two-derivative level and up to a 
certain level of accuracy, it should be possible to use an other effective theory, which 
is exactly supersymmetric and differs from the actual effective theory only by small effects 
related to the contribution of the heavy fields to supersymmetry breaking. It is then 
of general interest to understand more precisely under which conditions such a 
situation can arise and to develop a systematic procedure to construct the supersymmetric 
low-energy effective theory. 

The case of theories with global supersymmetry is well understood,
both for chiral \cite{ADS,IS} and vector multiplets \cite{PD,R,ADM}, but we will 
nevertheless review it in some detail. In this case one arrives very 
naturally at a simple procedure allowing to integrate out heavy superfields 
directly at the superspace level, and thus in a manifestly supersymmetric way. 
One particularly relevant situation where this procedure can be very 
usefully employed is that of supersymmetric Grand Unified Theories, 
with a high scale of gauge symmetry breaking yielding large masses for 
several fields \cite{PD,R}. The case of supergravity theories, 
on the other hand, seems to be less understood, and the main 
aim of this paper is to clarify how one should proceed in that case. 
For chiral multiplets, the question has been investigated some 
time ago in \cite{dA}, and some difficulties seem to appear, whereas
for vector multiplets the situation seems to be simpler \cite{ADM} 
(see also \cite{CJ,GRSDT,S}). We will however show that also in this case 
under suitable conditions one arrives at a simple prescription for integrating 
out heavy superfields in a manifestly supersymmetric way. 
This is particularly relevant in the context of the effective 
supergravity description of string models, where some of the moduli fields 
may be stabilized in a supersymmetric way with a large mass, like for example 
in the scenarios of \cite{GKP,KKLT}. There has been some debate on the 
circumstances in which it is justified to freeze such heavy moduli to constant 
values \cite{CF,dA2,AHK} (see also \cite{AHS}), and although this issue has recently 
been settled in \cite{GS, C1,C2}, it is important to know the 
procedure to integrate them out in general.

The crucial point to take into account when dealing with supersymmetric 
low-energy effective theories is that the usual expansion in number of derivatives 
does not preserve order by order supersymmetry. Any truncation on the number of 
derivatives spoils then supersymmetry, unless some other measure is taken. 
In fact, in a supersymmetric theory a restriction on the number of derivatives 
implies also a restriction on the numbers of fermions and auxiliary 
fields, due to the general form taken by supersymmetry transformations. 
More precisely, the only quantity that can be constrained consistently with 
supersymmetry is the total number $n$ of derivatives ($n_\partial$), fermion bilinears 
($n_\psi/2$) and auxiliary fields ($n_F$), defined as
\be
n = n_\partial + \frac12\, n_\psi + n_F \,.
\label{ndef}
\ee
One usually restricts from the beginning to supersymmetric theories with $n \le 2$. 
However, when integrating out heavy multiplets to define a low-energy effective 
theory valid below a certain mass scale $M$, infinitely many terms with 
arbitrarily large $n$ and coefficients suppressed by inverse powers 
of $M$ are in general generated. One may then decide to retain only 
those terms with $n \le 2$. But this truncation is justified solely when not only derivatives 
but also fermions and auxiliary fields are small in units of $M$. This means 
physically that the modes that are integrated out should not only be heavy, 
but also be stabilized in a way that approximately preserves supersymmetry,
with small values for the fermions and auxiliary fields, implying in particular
small mass splittings. The supersymmetric low-energy effective theory defined 
in this way, by truncating the total number of derivatives, fermion bilinears and 
auxiliary fields to $2$, is then different from the standard low-energy effective 
theory, obtained by truncating only the number of derivatives to $2$, and the two 
approximately coincide only in those regions of field space where fermions 
and auxiliary fields are small. One can summarize this reasoning by simply 
saying that a multiplet of fields can be integrated out in a supersymmetric 
way only if it has a large supersymmetric mass.

The main goal of this paper is to determine in a systematic way the effects
induced by integrating out heavy fields in the supersymmetric low-energy effective 
theory for light fields, under the assumptions defined above, that is, retaining only
terms with $n \le 2$. These corrections can be in general physically relevant and 
cannot just be discarded, except for very special situations. 
More concretely we will show how the heavy supermultiplets can be integrated out directly 
at the level of superfields, along the lines of \cite{PD,R,ADM} and thus in a very efficient way. 
The main result of our analysis is that heavy chiral superfields $\Phi^h$ and heavy vector 
superfields $V^x$ can be integrated out in an algebraic way by requiring respectively 
stationarity of the superpotential, $\partial_hW = 0$, and stationarity of the K\"ahler potential, 
$\partial_x K = 0$. We find that these equations have the same form in globally 
and locally supersymmetric theories. They substitute the conditions of stationarity of 
the potential with respect to the heavy fields $\xi^h$, $\partial_h V = 0$, that is 
normally used to define generic non-supersymmetric effective theories.

The basic reason why gravity does not affect the way in which one integrates 
out heavy fields at the leading order in the low-energy expansion is due to 
the fact that when requiring also gravity to be described at the two-derivative 
level, its couplings are essentially fixed. This is true in general, for any theory
with fields of spin $0$, $1/2$ and $1$, independently of whether it is 
supersymmetric or not. It can be understood through the following argument. 
A generic two-derivative theory without gravity is entirely parametrized by a 
potential $V$ and some wave-function factors $Z$ defining the kinetic terms,
which are functions of the fields. To get the effective theory at the two-derivative 
level, one can then integrate out the heavy fields $\xi^h$ by using as equations 
of motion $\partial_h V=0$ and completely neglecting space-time derivatives. 
As a matter of fact, this correctly determines not only the effective potential, 
but also the effective wave-function factors. The reason is that the
corrections to the equation $\partial_h V=0$ involve derivatives of the fields. 
Their effect can then be neglected in the wave function $Z$, since this 
would give terms with more than two derivatives in the action. It is easy 
to see that their effect can also be neglected in $V$. The reason for this is 
that only the leading linear effect can produce a term with two or less derivatives, 
but this term is proportional to $\partial_hV$ evaluated on the approximate 
solution and therefore vanishes. When switching on gravity, the potential and 
kinetic terms get covariantized in a unique way, and the only new allowed term 
is an Einstein-Hilbert kinetic term for gravity, multiplied by a function $\Omega$ 
of the fields. One can then repeat exactly the same reasoning as without gravity, treating 
$\Omega$ in a similar way as $Z$, and arrive again to the conclusion that one can 
use the simple equation $\partial_hV = 0$ to define the effective theory at the
two-derivative level. The case of supersymmetric theories is then just a special 
case of this. For heavy chiral multiplets $\Phi^h$, the K\"ahler potential $K$ plays 
a role similar to $Z$, whereas the superpotential $W$ corresponds essentially to $V$. 
For heavy vector multiplets $V^x$, it is instead the gauge kinetic function $H$ that 
plays the role of $Z$ and the K\"ahler potential $K$ that plays the role of $V$. 
In the case where such heavy chiral and vector superfields are stabilized in 
an approximately supersymmetric way, the analogs of the equation $\partial_h V = 0$ 
turn then out to be respectively $\partial_h W=0$ and $\partial_x K=0$. 
For exactly the same reasons as before, these equations allow to correctly compute 
not only the effective potential but also the wave function factors, and turn out to be 
valid also in the presence of gravity. The only assumption behind this is that gravity 
can be treated at the two-derivative level, and we shall see that this implies that the 
gravitino mass should be small.

The structure of the paper is as follows. In section 2 we discuss in 
detail the case of chiral multiplets in global supersymmetry. In section 3 we 
generalize the same analysis to the case of local supersymmetry, 
emphasizing the new restrictions coming from gravity.
In section 4 and 5 we then study more briefly the case of vector multiplets
in global and local supersymmetry respectively. Finally, in section 6 we 
discuss some implications of our results and present our conclusions.

\section{Chiral multiplets in global supersymmetry}\setcounter{equation}{0}

Let us consider first the simplest case of a globally supersymmetric theory with 
light chiral multiplets $\Phi^l$ and heavy chiral multiplets $\Phi^h$, denoted 
collectively by $\Phi^i$. Using the standard superspace formalism \cite{WB}, 
the total number $n$ of derivatives, fermion bilinears and auxiliary fields 
corresponds simply to half the number of $\theta^\alpha$ or $\bar \theta^{\dot \alpha}$ 
integrations ($n_{d\theta}$) plus half the number of supercovariant derivatives 
$D_\alpha$ or $\bar D_{\dot \alpha}$ ($n_{D}$): $n = (n_{d\theta} + n_{D})/2$. 
Requiring $n \le 2$ corresponds then 
to neglect any dependence on supercovariant derivatives in the action 
functional.\footnote{Note that $W$ must be chiral and could thus possibly 
depend only on $\bar D^2 \bar \Phi^i$, besides $\Phi^i$. But any term involving 
at least one $\bar D^2 \bar \Phi^i$ can be rewritten as a total $\bar D^2$ derivative, 
and thus reinterpreted as a correction to $K$.} As a consequence, the theory 
can be entirely parametrized in terms of a real K\"ahler potential 
$K = K(\Phi^i, \bar \Phi^{\bar \imath})$ and a holomorphic superpotential 
$W = W(\Phi^i)$. Note that in general this theory may itself already be a 
non-renormalizable effective theory valid only up to some energy scale $\Lambda$,
as long that this scale is larger than the mass scale $M$ of the heavy fields. The 
Lagrangian of such a theory is simply given by:
\bea\label{lag}
{\cal L} = \int \! d^4 \theta \, K (\Phi^i, \bar \Phi^{\bar \imath}) + \int \! d^2 \theta \, 
W (\Phi^i) + \int \! d^2 \bar \theta \, \bar W (\bar \Phi^{\bar \imath}) \,. 
\eea
The exact superfield equation of motion for $\Phi^h$ is obtained by first 
rewriting the first term in eq.~(\ref{lag}) as an $F$-term by making use of 
supercovariant derivatives, and then varying ${\cal L}$ with respect to the unconstrained 
chiral superfield $\Phi^h$. This yields:\footnote{Here and in the following we use the 
standard notation in which lower indices on functions denote derivatives with respect 
to their arguments.}
\bea
W_h - \frac 14 \bar D^2 K_h  = 0 \,.
\label{eqexact}
\eea

The presence of a large supersymmetric mass for $\Phi^h$ means that around the 
value $\Phi_0^h$ at which the superfield is stabilized, the superpotential $W$ 
has a large second derivative $W_{h h'}(\Phi_0^h)$ setting the mass scale 
$M$. This implies that the first term in eq.~(\ref{eqexact}) dominates over the second, and 
therefore $\Phi_0^h$ is approximately determined by the equation $W_h(\Phi^h_0)=0$. 
The departure $\Delta \Phi_0^h$ from the approximate solution can be computed 
by expanding eq.~(\ref{eqexact}) around $\Phi_0^h$, and at first order one finds that 
$\Delta \Phi_0^{h} \sim {\cal O} (D^2\,\Phi^l/M)$. It turns then out that this 
deviation can be completely neglected in our approximation, as 
the leading corrections that it would give to the effective action would have $n=3$. 
This statement is obvious for the terms coming from $K$, which gives terms with $n = 2$ 
in the absence of extra supercovariant derivatives. For the terms coming from $W$, which 
gives terms with $n = 1$ in the absence of extra supercovariant derivatives, this is on the other 
hand due to the fact that the leading correction is proportional to $W_h$, and therefore 
vanishes on the leading order solution. Summarizing, one can thus integrate out 
the superfields $\Phi^h$ by using the simple chiral superfield equation
\bea
W_h = 0 \,.
\label{eqchiral}
\eea
This equation determines in an algebraic way the heavy chiral superfields in terms 
of the light chiral superfields:
\bea
\Phi^h = \Phi_0^h(\Phi^l) \,.
\eea
The effective theory for the $\Phi^l$ is then obtained by plugging back this solution 
into $K$ and $W$. This yields:
\bea
\a\a K^{\rm eff} (\Phi^l,\bar \Phi^{\bar l}) = K(\Phi^l,\bar \Phi^{\bar l}, \Phi_0^h(\Phi^l), \bar \Phi_0^{\bar h}(\bar \Phi^l)) \,, \nn \\
\a\a W^{\rm eff}(\Phi^l) = W(\Phi^l, \Phi_0^h(\Phi^l)) \,.
\eea

It is instructive to rederive these results by using component fields. The Lagrangian 
takes then the usual form ${\cal L} = T - V$, where the kinetic term reads
\bea
T = - K_{i \bar \jmath} \big(\partial_\mu \phi^i \partial^\mu \bar \phi^{\bar \jmath} 
+ i \bar \psi^{\bar \jmath} \bar \sigma^\mu D_\mu \psi^i \big) \,,
\eea
with $D_\mu \psi^i = \partial_\mu \psi^i + K^i_{mn} \partial_\mu \phi^m \psi^n$, 
and the potential is given by
\bea
V &=& - W_i F^i - \bar W_{\bar \jmath} \bar F^{\bar \jmath} 
+ \frac 12 W_{ij} \psi^i \psi^j + \frac 12 \bar W_{\bar \imath \bar \jmath} \bar \psi^{\bar \imath} \bar \psi^{\bar \jmath} \nn \\
&\;& -\, K_{i \bar \jmath} F^i \bar F^{\bar \jmath}
+ \frac 12 K_{i \bar \jmath \bar k} F^i \bar \psi^{\bar \jmath} \bar \psi^{\bar k} 
+ \frac 12 K_{\bar \jmath m n} F^{\bar \jmath} \psi^m \psi^n
- \frac 14 K_{i \bar \jmath p \bar q} \psi^i \psi^p \bar \psi^{\bar \jmath} \bar \psi^{\bar q} \,.
\eea
Recall also that the auxiliary fields are actually determined by their algebraic equations of motion, 
and are given by:
\be
F^i = - K^{i \bar \jmath} \Big(\bar W_{\bar \jmath} - \frac 12 K_{\bar \jmath m n} \psi^m \psi^n\Big) \,.
\label{Fi}
\ee
We can now derive the exact equations of motion of $F^h$, $\psi^h_{\alpha}$ and $\phi^h$.
These correspond to the $\theta^0$, $\theta^\alpha$ and $\theta^2$ components of 
(\ref{eqexact}) and determine respectively the values of the auxiliary fields $F^h$, the 
wave equation for $\psi_\alpha^h$ and the wave equation for $\phi^h$. One finds, 
without needing to use eq.~(\ref{Fi}), the following equations:
\bea
\a\a W_h + K_{h \bar \jmath} \bar F^{\bar \jmath} 
- \frac 12 K_{h \bar \imath \bar \jmath} \bar \psi^{\bar \imath} \bar \psi^{\bar \jmath} = 0 \,, 
\label{eqexact1} \\
\a\a W_{h i} \psi^i_\alpha + K_{h i \bar \jmath} \psi^i_\alpha \bar F^{\bar \jmath} 
- \frac 12 K_{h i \bar \jmath \bar k} \psi^i_\alpha \bar \psi^{\bar \jmath} \bar \psi^{\bar k} 
+  i K_{i \bar \jmath} \sigma^\mu D_\mu \bar \psi^{\bar \jmath} = 0 \,, 
\label{eqexact2} \\
\a\a W_{h i} F^i - \frac 12 W_{h ij} \psi^i \psi^j + K_{h i \bar \jmath} F^i F^{\bar \jmath}
- \frac 12 K_{h i \bar \jmath \bar k} F^i \bar \psi^{\bar \jmath} \bar \psi^{\bar k} 
- \frac 12 K_{h \bar \jmath m n} F^{\bar \jmath} \psi^m \psi^n \nn \\
\a\a \,+\, \frac 14 K_{h i \bar \jmath p \bar q} \psi^i \psi^p \bar \psi^{\bar \jmath} \bar \psi^{\bar q} 
+ K_{h \bar \jmath} \Box \bar \phi^{\bar \jmath} 
+ K_{h \bar \jmath \bar k} \partial_\mu \bar \phi^{\bar \jmath} \partial^\mu \bar \phi^{\bar k} = 0 
\label{eqexact3}\,.
\eea
Under supersymmetry transformations, these equations get mapped into each other 
and remain thus satisfied.

In the situation in which the fields $\phi^h$ and $\psi^h_\alpha$ have a large supersymmetric mass
$M$, there must be a quadratic term in $W$ leading to a second derivative $W_{h h'}$ of order
$M$. The equations of motion (\ref{eqexact2}) and (\ref{eqexact3}) for $\psi_\alpha^h$ and $\phi^h$
are then dominated by the first terms,  which involve second derivatives of $W$. 
Similarly, in the equation of motion (\ref{eqexact1}) for $F^h$, the first term is 
expected to dominate, since the other two do not involve $W$ at all. In the brutal limit in which one 
takes $M\rightarrow \infty$ one would find that $\phi^h$ is determined by the condition 
$W_h (\phi^h) = 0$ whereas $\psi_\alpha^h$ and $F^h$ vanish. However, this brutal approximation 
does not preserve supersymmetry. One needs therefore to look at the subleading terms and check 
which ones should be kept in order to get a set of equations that is supersymmetric. 
The appropriate criterion to do so is related to the counting of the total number $n$ of derivatives, 
fermion bilinears and auxiliary fields. Indeed, in order to obtain an effective theory with $n \le 2$, 
each of the equations used to integrate out the heavy fields in terms of the light ones should involve 
terms with the same minimal value of $n$. Looking at eqs.~(\ref{eqexact1})--(\ref{eqexact3}), it is easy to see  
that the terms depending on $W$ have a value of $n$ that is one unit less than the terms 
depending on $K$ and are therefore the dominant ones. One may then drop all the terms 
involving $K$ and find the following set of approximate equations:
\bea
\a\a W_h = 0 \,, \label{eqchiral1} \\[1mm]
\a\a W_{h i} \psi^i_\alpha = 0 \,, \label{eqchiral2} \\
\a\a W_{h i} F^i - \frac 12 W_{h ij} \psi^i \psi^j = 0 \,. \label{eqchiral3} 
\eea
It is easy to check that these are now exactly supersymmetric. More precisely, under supersymmetry 
transformations each equation transforms into a combination of its space-time derivative and one of the 
other equations. These equations are in fact the non-trivial components of a chiral superfield equation, 
which is nothing but eq.~(\ref{eqchiral}). The first of them is now understood as determining $\phi^h$,
the second $\psi_\alpha^h$ and the third $F^h$. 
The bottom line is that the appropriate equation to be used to integrate out the scalar fields is indeed 
the naive one, whereas for the fermion and auxiliary fields supersymmetry forces us to keep some 
subleading terms suppressed by the mass scale $M$. 

Let us finally analyze in a more physical way the situation by disentangling the physical fields 
$\phi^i$ and $\psi_\alpha^i$ from the auxiliary fields $F^i$. The potential $V$ for the physical
fields can be computed by eliminating the auxiliary fields through their equations of motion (\ref{Fi}).
The mass matrices of the physical fluctuation fields $\varphi^i$ and $\chi_\alpha^i$ around a generic 
point in field space, with arbitrary value of the scalar fields but vanishing value for the fermion fields, 
are then determined by the second 
derivatives of such a potential. In the limit where all the $F^i$ are small, the only terms that survive in 
these mass matrices correspond to a common supersymmetric mass coming from a quadratic term 
in the superpotential: 
\be
{\cal L}_{\rm mass} = -\, W_{i p} K^{p \bar q} \bar W_{\bar q \bar \jmath} \, \varphi^i \bar \varphi^{\bar \jmath}
- \frac 12 \big(W_{ij} \chi^i \chi^j + \bar W_{\bar \imath \bar \jmath} \bar \chi^{\bar \imath} \bar \chi^{\bar \jmath}\big) \,.
\ee
The supersymmetric mass matrix is thus described by the complex matrix $W_{ij}$. The blocks
corresponding to the heavy fields, the light fields and their mixing are:
\be
M_{h h'} = W_{h h'} \,,\;\;
m_{l l'} = W_{l l'} \,,\;\;
\mu_{h l} = W_{h l} \,.
\label{masses}
\ee
The other relevant parameters are the cubic couplings in the superpotential involving heavy chiral 
multiplets, namely:
\be
\eta_{h l l'} = W_{h l l'} \,,\;\; \delta_{h h' l} = W_{h h' l} \,,\;\; \xi_{h h' h''} = W_{h h' h''} \,.
\ee
The physical masses are finally obtained by rescaling the fields in order to canonically normalize their kinetic terms,
which involve the K\"ahler metric $K_{i \bar \jmath}$. This can be done by using the vielbeins of the K\"ahler 
manifold spanned by the scalar fields. In the end, one finds that the mass eigenstates are linear combinations of 
the chiral multiplets, consisting each of a scalar, a pseudoscalar and a fermion with equal mass.

Using the identifications (\ref{masses}), we can now spell out more concretely the content of the three components 
(\ref{eqchiral1})--(\ref{eqchiral3}) of the superfield equation (\ref{eqchiral}). The first equation states that 
the $\phi^h$ must adjust to values compatible with the assumption that all the $\psi_\alpha^h$ and $F^h$ 
vanish in first approximation:
\be
\phi_0^h(\phi^l) : \text{solution of}\; W_h (\phi^l, \phi_0^h) = 0\,.
\label{phih0}
\ee
The second equation tells us instead that the $\psi_\alpha^h$ are not exactly zero but proportional to the 
$\psi^l$, through a coefficient given by the ratio between the mass mixing $\mu$ between light and heavy 
fields and the mass $M$ of the heavy fields:
\bea
\psi_{0\alpha}^h (\phi^l,\psi_\alpha^l) = - (M_0^{-1} \! \mu_0)^h_{\;\,l} (\phi^l)\, \psi_\alpha^l \,.
\label{psih0}
\eea
Finally, the third equation implies that the $F^h$ are not exactly zero either, but 
proportional to the $F^l$, plus some terms quadratic in the $\psi_\alpha^l$, again
through coefficients involving the ratio between $\mu$ and $M$. Schematically one finds:
\bea
F_0^h(\phi^l,\psi^l_\alpha) &=& - (M_0^{-1} \! \mu_0)^h_{\;\,l} (\phi^l) F^l \nn \\
&\;& \,- \frac 12 \big(M_0^{-1} \eta_0 - 2 M_0^{-2} \! \mu_0 \delta_0 + M_0^{-3} \! \mu_0^2 \xi_0\big)^h_{\; l l'} (\phi^l)\psi^l \psi^{l'} \,.
\label{Fh0}
\eea
In summary, we see that this procedure automatically keeps track of the fact that the heavy superfields 
have small but yet non-vanishing fermion and auxiliary field components. The effects of these suppressed 
components are in general relevant and cannot be neglected. The final result 
is then a supersymmetric effective theory that is accurate at leading order in $\partial^\mu/M$, $\psi^i/M^{3/2}$
and $F^i/M^2$, but a priori not limited to small $\phi^i/M$. 

We have checked in a variety of examples that the supersymmetric effective theory defined by the superfield 
equation $W_h = 0$ and the standard effective theory defined by the ordinary equation $V_h = 0$ do indeed
approximately coincide under the above assumptions. Focusing for concreteness on the scalar potential,
the region in the space of scalar fields $\phi^l$ where the two theories match is defined by the following two 
conditions:\footnote{Note that in general the whole supersymmetric mass matrix, including all the blocks 
$m$, $\mu$ and $M$, is field dependent. One has therefore to make sure that not only $m$ but also the 
mixing term $\mu$ stay small compared to $M$ (see also \cite{L} regarding this point). One can however 
focus on the supersymmetric part of the mass matrix, since the $F^i$ are independently assumed to be small.}
\be
m(\phi^l),\mu(\phi^l) \ll M \,,\;\; F^l(\phi^l),F^h(\phi^l) \ll M^2 \,.
\ee
In the simple case of theories with $K$ and $W$ at most quadratic in the fields, one can check 
analytically that the difference between the two effective potentials is proportional to some positive
power of $F^l(\phi^l)$. In more general theories, instead, one can perform only a point-by-point 
numerical check. 

One can redo the same analysis in the presence of additional light vector superfields $V^a$,
which can always be viewed as associated to local gauge symmetries.\footnote{If $V^a$ is a 
general real vector multiplet, one can reformulate the theory by adding a would-be Goldstone 
chiral multiplet and a gauge symmetry, in such a way to go back to the situation in which one has
gauge vector superfields plus charged chiral superfields.} 
The only relevant difference is that the K\"ahler potential $K$ can now also depend on these 
extra fields. The counting of the total number $n$ of derivatives, fermion bilinears and auxiliary 
fields gets then modified.\footnote{We thank the authors of ref.~\cite{GS} for drawing our attention 
on this issue and sharing with us some preliminary results that will appear in \cite{GS2}.} 
The reason is that the vector multiplets $V^a$ have dimension $0$ rather than $1$ as the chiral 
multiplets. We choose to keep the definition (\ref{ndef}) of $n$ as a generalized number of derivatives 
and count $A_\mu^a$, $\lambda^a$ and $D^a$ with $n=0,1/2$ and $1$. The additional components 
$c^a, \chi^a$ and $N^a$ arising in non-Wess-Zumino gauges must then be assigned $n=-1,-1/2$ and 
$0$ to preserve supersymmetry. This counting guarantees that the minimal Lagrangian has $n \le 2$, 
as before, but has the disadvantage of not preserving gauge invariance. The important novelty is that 
there can now be terms in $K$ with $n=0$ and $n=1$, even in the Wess-Zumino gauge.
This raises then the question of whether one should in this case keep subleading terms with $2$ 
and $4$ supercovariant derivatives in the solution of the superfield equations of motion of the heavy 
chiral multiplets, besides the approximate solution obtained by solving the truncated equation 
$W_h = 0$. It turns out that this is again not necessary, but for a slightly less trivial reason than 
before. Indeed these subleading terms induce interactions with additional supercovariant derivatives 
acting on the light vector multiplets and/or the light chiral multiplets. The former can give 
terms with $n=2$, whereas the latter can give only terms with $n=3,4$. But these different kinds 
of terms are related to each other by superspace gauge transformations. Since terms with $n=3,4$
must certainly be neglected, one is forced to neglect the new terms with $n = 2$ as well. 
One may then wonder whether and why this is justified. The answer to this question comes 
from the observation that although these terms have $n=2$, they come with a 
coefficient that contains one or two powers of the heavy chiral multiplet mass $M$ in the denominator,
and thus one or two powers in the numerator of some other parameter $m$ with the dimension of a mass 
that is related to the couplings between $V^a$ and $\Phi^h$. But it is clear that this parameter must be 
of the order of the mass of the vector multiplets $V^a$. As a result, although the new terms can 
arise at the allowed order in the $n$ counting, they are further suppressed by at least one or two 
powers of the ratio $m/M$ between light and heavy masses, and can thus be neglected. We have 
checked in a few non-trivial examples that this is indeed what happens. Focusing on the potential, 
for instance, one finds terms of order $(m/M) F D$ and $(m/M)^2 D^2$, which can be neglected.

\section{Chiral multiplets in local supersymmetry}\setcounter{equation}{0}

Let us consider next the case of a locally supersymmetric theory with light chiral multiplets $\Phi^l$ 
and heavy chiral multiplets $\Phi^h$, denoted collectively by $\Phi^i$, as well as the gravitational 
multiplet. It will be convenient for our purposes to use the superconformal superspace formalism
\cite{CONF1,CONF2,CONF3}, 
where the gravitational sector is described by a conformal gravitational multiplet $G$, containing 
the graviton, the gravitino and two vector auxiliary fields, and a chiral compensator multiplet $\Phi$,
containing one complex scalar field, one fermion field and one scalar auxiliary field. In this formalism, 
the supersymmetry transformations, the tensor calculus and the superspace structure are very similar 
to those of rigid supersymmetry, in the sense that they are deformed through terms depending on 
the fields of $G$ but not those of $\Phi$. The superconformal group can then be reduced to the 
super-Poincar\'e group by gauge fixing the additional symmetries. More precisely, one can fix the 
scalar and fermionic components of $\Phi$ and also get rid of one of the two vector auxiliary fields 
of $G$. In this way, one gets back the ordinary formulation of supergravity \cite{SUGRA1A,SUGRA1B}, with ordinary 
supersymmetry transformations emerging as a combination of supersymmetry and extra conformal transformations 
preserving the superconformal gauge choice. Alternatively, one may also choose to keep all 
the fields together with the additional superconformal symmetries, and this proves to be useful in 
some instances. In this formalism, the total number $n$ of derivatives, fermion bilinears and scalar 
auxiliary fields corresponds again to half the number of Berezin integrals plus half the number 
of supercovariant derivatives, as far as the matter and compensator chiral multiplets are concerned. 
On top of that, however, one has also to consider the contribution from the gravitational multiplet, which 
can also bring derivatives and fermions bilinears, but no scalar auxiliary fields. Requiring $n \le 2$ 
amounts then to neglect supercovariant derivatives and in addition to work at the two-derivative/four-fermion 
level in the gravitational sector. The theory can again be parametrized in terms of a real K\"ahler 
potential $K=K(\Phi^i,\bar \Phi^{\bar \imath})$ and a holomorphic superpotential $W = W(\Phi^i)$. 
Notice that such a theory is unavoidably itself only a low-energy effective theory valid below the 
Planck scale $M_{\rm P}$, and the mass scale $M$ of the heavy fields should be smaller than this 
scale. Setting from now on $M_{\rm P} = 1$, the Lagrangian can be written in the following form:
\be\label{lag2}
{\cal L} = \int \! d^4 \theta \, \Big(\!-\! 3 \, e^{-K/3}\Big) \bar \Phi \Phi
+ \int \, d^2 \theta \, W\, \Phi^3 + \int \, d^2 \bar \theta \, \bar W\, \bar \Phi^3\,. 
\ee
The superspace integrals in the above expression generate components that 
involve terms depending on the fields of the gravitational multiplet $G$. However, these extra
terms are uniquely determined, and we will thus not keep track of them. 
By doing so, one can then manipulate superspace quantities exactly as 
in the rigid case. In particular, the exact superfield equations of motion for the heavy superfields 
$\Phi^h$ are given by:
\bea
W_h - \frac 14 \bar D^2 \Big(K_h e^{-K/3} \Phi^\dagger \Big) \Phi^{-2} = 0 \,.
\label{eq1}
\eea

We assume that as before the presence of a large supersymmetric mass 
means that around the value $\Phi_0^h$ at which the heavy superfields $\Phi^h$ are stabilized, the 
superpotential  $W$ has a large second derivative $W_{h h'}(\Phi_0^h)$ setting the 
mass scale $M$. The equations of motion are then dominated by the first term, and read in first approximation
$W_h(\Phi_0^h) = 0$. The leading deviation $\Delta \Phi_0^h$ from this approximate solution
can be evaluated by expanding eq.~(\ref{eq1}) around $\Phi_0^h$, and one finds that 
$\Delta \Phi_0^h \sim {\cal O}(D^2\Phi^l/M, D^2\Phi/M)$. This deviation can be neglected,
since it would as before give corrections with $n>2$.\footnote{A similar reasoning has also 
been used in \cite{C1,C2} in the special case of effective theories describing string models 
with fluxes.} The heavy chiral superfields can thus 
be integrated out by using the same simple chiral superfield equation as in the rigid case, namely
\be
W_h = 0 \,.
\label{eq1app}
\ee
As before, the solution of this equation determines the heavy chiral fields in terms of the light chiral fields:
\bea
\Phi^h = \Phi_0^h(\Phi^l) \,.
\eea
The effective theory for the $\Phi^l$ is then obtained by plugging back this solution 
into $K$ and $W$. This yields:
\bea
\a\a K^{\rm eff} (\Phi^l,\bar \Phi^{\bar l}) = K(\Phi^l,\bar \Phi^{\bar l}, \Phi_0^h(\Phi^l), \bar \Phi_0^{\bar h}(\bar \Phi^l)) \,, \nn \\
\a\a W^{\rm eff}(\Phi^l) = W(\Phi^l, \Phi_0^h(\Phi^l)) \,.
\eea
Notice now that the original theory involving all the fields has a K\"ahler symmetry 
acting as $(\Phi, K, W) \to (\Phi \,e^{X/3}, K + X + X^\dagger, W e^{-X})$, where 
$X(\Phi^i)$ is an arbitrary holomorphic function of the matter chiral superfields. 
On the other hand, the superfield equation (\ref{eq1app}) defining the 
effective theory is not manifestly invariant under such a transformation 
for generic $X$. More precisely, it is invariant if $X$ depends only on the $\Phi^l$, 
corresponding to K\"ahler transformations within the effective theory. But it is 
not invariant if $X$ depends also on the $\Phi^h$. The reason for this is that we 
have assumed in our derivation that the large mass scale $M$ of the heavy fields 
is associated only with a large quadratic term in $W$, and no large term in $K$. 
This selects a restricted subclass of K\"ahler gauges, which is particularly well-suited 
to work out the effective theory. 

One may wonder at this point whether it is really justified to neglect supercovariant 
derivatives acting on the compensator, and try to see what is the outcome when one keeps such 
terms and neglects only those where supercovariant derivatives act on the other chiral 
superfields. Proceeding in this way, eq.~(\ref{eq1}) would not reduce to eq.~(\ref{eq1app}), 
but rather to
\bea
W_h - \frac 14 \Phi^{-2} K_h e^{-K/3}   \bar D^2 \Phi^\dagger  = 0 \,.
\label{eq2}
\eea
In order to get rid of the dependence on the compensator, one can now use the exact 
superfield equation of motion of $\Phi$, without doing any superconformal gauge fixing, 
with the understanding that this equation will be partly related by the extra conformal 
symmetries to the wave equation of some modes of the fields in $G$. From the Lagrangian 
(\ref{lag2}), one finds that this equation of motion is given by
\bea\label{eq3}
W  + \frac 14 \bar D^2 \Big(e^{-K/3} \Phi^\dagger \Big) \Phi^{-2}= 0 \,.
\eea
For the same reasons as before, all the terms involving supercovariant derivatives
acting on $K$ can certainly be neglected. However, one should keep the terms where 
the supercovariant derivatives act on the compensator. Eq.~(\ref{eq3}) 
becomes then
\bea
- \frac 14 \bar D^2 \Phi^\dagger \Phi^{-2} = e^{K/3} W \,.
\label{eqcomp}
\eea
Plugging this relation back into eq.~(\ref{eq2}) allows finally to eliminate 
completely the dependence on the compensator, and the final equation simply 
reads:
\bea
W_h + K_h W = 0 \,.
\label{DHW}
\eea
This equation can also be derived in a more direct way by choosing from the beginning 
a K\"ahler gauge defined by $X = \ln W$. In this way one does not need to use the 
compensator equation of motion, but the derivation still implicitly assumes that 
$W \neq 0$ and $\bar D^2 \Phi^\dagger \neq 0$. 

Notice that eq.~(\ref{DHW}) reduces to the equation $W_h=0$ in the rigid limit, and is 
moreover manifestly invariant under K\"ahler transformations. However, a closer look  
shows that it cannot possibly be the correct equation. An obvious problem is that 
it is a vector and not a chiral superfield equation. This means that it 
cannot be solved by just setting the $\Phi^h$ to some functions of the $\Phi^l$, 
due to the fact that there are more component equations than component fields. 
On the other hand, the original 
exact equation of motion (\ref{eq1}) for $\Phi^h$ is chiral, and it is by dropping only part of the terms 
involving supercovariant derivatives that one arrives at an equation which is no longer chiral. 
Thus, the new equation must somehow also be approximately chiral, 
meaning that only its chiral components should really be significant, the non-chiral ones being 
approximately satisfied in an automatic way. This means that the equation $W_h + K_h W= 0$ 
cannot be used as an exact equation to define a manifestly supersymmetric approximate version 
of the low-energy effective field theory, and that the appropriate equation should instead be $W_h=0$,
as already argued. Through this argument, we have moreover learned that neglecting terms involving 
supercovariant derivatives acting on the compensator amounts to neglect $W$ compared to $M$, 
i.e. to have approximately
\be
W \simeq 0 \,.
\ee
This equation should however not be imposed as an exact superfield equation as it comes from a 
reasoning on the compensator superfield $\Phi$, for which most of the components can be gauged 
away. More precisely, in the formulation where the superconformal symmetry is gauge-fixed to the 
super-Poincar\'e symmetry, only the lowest component of this equation, corresponding to the 
equation coming from the auxiliary field of the compensator, should be considered.
Finally, it should also be emphasized that although $W$ must be neglected in the 
equation that is used to integrate out the $\Phi^h$, one should a priori not neglect 
terms involving $W$ in the Lagrangian where the solution for the $\Phi^h$ is substituted 
to obtained the effective theory for the $\Phi^l$.

The crucial point behind this extra difficulty that one encounters in the gravitational case 
is that space-time derivatives and supersymmetry-breaking auxiliary fields must be small
also in the gravitational sector. This brings up a new condition that needs to be fulfilled 
in order to be in the situation in which an approximate two-derivative supersymmetric 
low-energy effective theory is expected to exist: the compensator auxiliary field $F$ should 
be much smaller than $M$:
\be
F \ll M \,.
\ee
Once all the other auxiliary fields $F^i$ are also assumed to be small, $F^i \ll M^2$, 
this condition implies that: a) the gravitino mass (and therefore $W$) is small, $m_{3/2} \ll M$, 
as $m_{3/2}$ is a linear combination of $F$ and the $F^i$, and b) the cosmological 
constant is small, $V \ll M^2$, as $V$ is a quadratic combination of $F$ and the $F^i$.
The reason for requiring $F$ to be small is then twofold. On one hand, it can represent 
a supersymmetry-breaking effect in a flat background space-time, and should then be small
in order not to induce too large mass splittings. On the other hand, it can also represent
a supersymmetry-preserving cosmological constant implying a curved background 
space-time, and should then be small in order to allow a two-derivative approximation 
for the graviton, which is justified only for small curvature. It should be emphasized that 
this further condition $F \ll M$ (or equivalently $m_{3/2} \ll M$) can in general not be achieved in a 
natural way, but must instead be implemented through an adjustment of parameters in the 
Lagrangian. Notice however that for phenomenological applications it is anyhow necessary 
to eventually tune this cosmological constant to a yet smaller value in the low-energy 
effective theory. This step does therefore not represent a really severe restriction. 
Nevertheless, it is not possible to define a locally supersymmetric two-derivative low-energy 
effective theory below $M$ without making sure that this condition is satisfied.

One can derive the same results using component fields in the ordinary formulation of 
supergravity. This is recovered after gauge-fixing the additional conformal symmetries 
by setting one of the vector auxiliary fields of $G$ to zero and the scalar and fermionic 
components of $\Phi$ to some reference values. In order to work directly in the Einstein 
frame, we shall set the compensator scalar to $\phi = e^{K/6}$, which is a function of the 
other scalar fields $\phi^i$. Similarly, the various field redefinitions that are usually performed  
to simplify the Lagrangian in the fermionic sector can be achieved by setting the compensator 
fermion $\psi$ to a suitable linear combination of the other fermions $\psi_\alpha^i$.
Finally, it is convenient to parametrize the compensator auxiliary field as $F = e^{K/6} U$. 
In this way $U$ corresponds to the usual scalar auxiliary field of supergravity. 
After this gauge fixing, only a combination of the original supersymmetry with the additional
conformal symmetries survives, which correspond to the ordinary Poincar\'e supersymmetry 
transformations. For simplicity, we shall not keep track of the fermions and focus only on the 
bosonic fields. Moreover, we will also discard the dependence on the gravitational fields, 
except for the scalar auxiliary field $U$ originating from the compensator. 
In these approximations, the Lagrangian has the usual form ${\cal L} = T - V$, 
with a kinetic term that is the same as in global supersymmetry,\footnote{Note that the K\"ahler 
covariant derivative emerges only after taking into account the couplings to the vector auxiliary 
field that remains in the gravitational multiplet after superconformal gauge fixing.} 
\be
T = - K_{i \bar  \jmath} \partial_\mu \phi^i \partial^\mu \bar \phi^{\bar \jmath} \,,
\ee
and a potential taking the following form:
\bea
V &=&  - \, W_i F^i  e^{K/2} - \bar W_{\bar \jmath} \bar F^{\bar \jmath}  e^{K/2}
- 3\, W  U  e^{K/2} - 3\, \bar W \bar U e^{K/2} \nn \\
&\;& -\, \Big(K_{i \bar  \jmath} - \frac 13 K_i K_{\bar \jmath} \Big)F^i \bar F^{\bar  \jmath} 
- K_i F^i \bar U - K_{\bar \jmath} \, \bar F^{\bar \jmath} U + 3 \, U \bar U \,.
\eea
Recall also that the auxiliary fields $F^i$ and $U$ are determined by their algebraic equations of motion,
which give:
\bea
F^i &=& - K^{i \bar \jmath} \big(\bar W_{\bar \jmath}  \hspace{-1pt}+ \hspace{-1pt} K_{\bar \jmath} \bar W \big) e^{K/2} \,, \label{Fieq}\\
U &=& \Big(1  \hspace{-1pt}- \hspace{-1pt} \frac 13 K_i K^{i \bar \jmath} K_{\bar \jmath} \Big) \bar W e^{K/2} 
- \frac 13 K_i K^{i \bar \jmath} \bar W_{\bar \jmath} e^{K/2} \label{Ueq} \,.
\eea
From these equations it follows that:
\bea
W e^{K/2} &=& \bar U - \frac 13 K_{\bar \jmath} \bar F^{\bar \jmath} \,, \label{Weq}\\
W_i e^{K/2} &=& - \Big(K_{i \bar \jmath} - \frac 13 K_i K_{\bar \jmath} \Big) \bar F^{\bar  \jmath} - K_i \bar U  \label{Wieq}\,.
\eea
We can now derive the equations of motion of $F^h$ and $\phi^h$. These correspond to the 
$\theta^0$ and $\theta^2$ components of the exact equations of motion after performing the 
superconformal gauge fixing on the compensator. One finds:
\bea\label{uno}
\a\a W_h e^{K/2} + \Big(K_{h \bar \jmath} - \frac 13 K_h K_{\bar \jmath} \Big)\bar F^{\bar  \jmath} + K_h \bar U = 0 \,, \\
\a\a  W_{hi} F^i e^{K/2} + \Big(K_{h i \bar \jmath} - \frac 13 \big(K_i K_{h \bar \jmath} + K_{\bar \jmath} K_{h i}\big) 
- K_h K_{i \bar \jmath} + \frac 13 K_h K_i K_{\bar \jmath} \Big) F^i \bar F^{\bar \jmath} \nn \\\label{dos}
\a\a \, + \Big(K_{h i} - K_h K_i \Big) F^i \bar U - 2 K_{h \bar \jmath} \bar F^{\bar \jmath} U
+ K_{h \bar \jmath} \Box \bar \phi^{\bar \jmath} 
+ K_{h \bar \jmath \bar k} \partial_\mu \bar \phi^{\bar \jmath} \partial^\mu \bar \phi^{\bar k} = 0 \,.
\eea
In order to arrive at this last equation, we have used the relations (\ref{Wieq}) and (\ref{Weq}) that follow 
from eqs.~(\ref{Fieq}) and (\ref{Ueq}).

To define an approximate low-energy effective theory, we can now neglect in each of these equations
those terms which are subleading in the counting of the total number $n$ of derivatives and auxiliary fields.
In this way we get:
\bea
\a\a W_h = 0 \,, \label{eq1app1}\\
\a\a W_{hi} F^i = 0 \label{eq1app3}\,.
\eea
We recognize now that these equations correspond indeed to the $\theta^0$ and $\theta^2$ components of the 
superfield equation (\ref{eq1app}) obtained in the superfield approach.

To check the effect of the compensator auxiliary field $F$, one may re-do the same analysis without 
considering it as an auxiliary field but rather as an ordinary scalar field. This can be easily done by first 
eliminating the field $U$ from the two equations for $\phi^h$ and $F^h$ by using its equation of motion
$U =  \bar W e^{K/2} + 1/3 K_i F^i$. Using also eq.~(\ref{Fieq}), and dropping then in (\ref{uno}) and (\ref{dos}) 
only terms that are subleading in the total number of derivatives and matter auxiliary fields, one would find 
the following two equations:
\bea
\a\a \big(W_h + K_h W \big) e^{K/2} = 0 \,, \\
\a\a \big(W_{hi} + K_{hi} W + K_h W_i \big) F^i e^{K/2} - 2\,  K_{h \bar \jmath}  \bar W \bar F^{\bar \jmath} e^{K/2}  = 0 \,.\label{dosd}
\eea
These should correspond to the $\theta^0$ and $\theta^2$ components of eq.~(\ref{DHW}). As a matter of fact, 
this is indeed the case if one discards the last term in eq.~(\ref{dosd}). This is related to the fact that 
(\ref{DHW}) is a vector superfield equation which is only approximately chiral and has, as already argued, 
also some $\bar \theta^2$ and $\theta^2 \bar \theta^2$ components that must 
somehow be automatically satisfied within our approximations. Its $\bar \theta^2$ component, in particular, 
implies that the quantity $K_{h \bar \jmath}  \bar W \bar F^{\bar \jmath}$ should be discarded. Under this 
assumption, the above equations correspond then indeed to the chiral components of eq.~(\ref{DHW}). 
As already argued, this equation cannot be taken as an exact superfield equation, and this shows up here 
through the fact that the above set of component equations is not preserved by supersymmetry transformations.

Let us finally summarize the situation by disentangling the physical fields 
$\phi^i$ from the auxiliary fields $F^i$ and $U$, discarding again the fermions $\psi^i$ for simplicity. 
The scalar potential $V$ for the $\phi^i$ can be computed by eliminating all the auxiliary fields through their 
equations of motion. The mass matrices of the physical fluctuation fields $\varphi^i$ 
around a generic point in field space are then determined by the second derivatives of this potential. 
In the limit where all the $F^i$ as well as $U$ are small, the only terms that survive in these mass 
matrices correspond to the usual supersymmetric mass coming from a quadratic term in the superpotential:
\be
{\cal L}_{\rm mass} = -\, e^K W_{i p} K^{p \bar q} \bar W_{\bar q \bar \jmath} \, \varphi^i \bar \varphi^{\bar \jmath} \,.
\ee
The supersymmetric mass matrix is thus described by the complex matrix $e^{K/2} W_{ij}$, and the three blocks 
corresponding to the heavy fields, the light fields and the mixing between the two are given by:
\be
M_{h h'} = e^{K/2} W_{h h'} \,,\;\;
m_{l l'} = e^{K/2} W_{l l'} \,,\;\;
\mu_{h l} = e^{K/2} W_{h l} \,.
\ee
The physical masses are obtained by rescaling the fields in order to canonically normalize the kinetic terms.
The mass eigenstates consistent of pairs of degenerate scalars and pseudoscalars with equal mass.

In this approach too, it is instructive to check more explicitly the role of the compensator auxiliary 
field by computing the masses in the limit where only the $F^i$ are neglected and $U$ is kept. 
In this way one finds that the mass terms become:
\bea
{\cal L}_{\rm mass} &=& - \big(N_{i p} K^{p \bar q} \bar N_{\bar q \bar \jmath} - 2 K_{i \bar \jmath} |U|^2 \big) 
\varphi^i \bar \varphi^{\bar \jmath} + \frac 12 N_{ij} U \varphi^i \varphi^j 
+ \frac 12 \bar N_{\bar \imath \bar \jmath} \bar U \bar \varphi^{\bar \imath} \bar \varphi^{\bar \jmath} \,,
\label{massesgra}
\eea
where
\be
N_{ij} = e^{K/2} W_{ij} + (K_{ij} - K_i K_j) \bar U \,.
\ee
The physical masses are then no-longer degenerate in pairs, but display now a splitting between 
scalars and pseudoscalars, of the order of the off diagonal elements $N_{ij} U$ in eq.~(\ref{massesgra}).
If supersymmetry is unbroken and the background geometry is AdS, the mass splittings coincide with those 
required by the supersymmetry algebra in AdS space. $U$ represents then a curvature scale and more 
precisely the inverse of the radius $L$ of AdS. In this case one must require that the Compton wave length 
$1/M$ of the heavy fields should be much smaller that this curvature length $L$, in order to be able to integrate out 
these states in the small curvature approximation. This implies in particular $U \ll M$. If on the other hand supersymmetry 
is broken and the background geometry is Minkowski, the mass splittings represent a soft supersymmetry 
breaking effect. $U$ corresponds then to an effective supersymmetry breaking scale. In this case one must 
require that the square mass $M^2$ of the heavy fields should be much larger than the mass splittings of 
order $M U$ and $U^2$ arising in eq.~(\ref{massesgra}). This implies again $U \ll M$.
Notice finally that if the condition $U \ll M$ is not satisfied, it is impossible for any light chiral multiplets
to heave both its scalar and pseudo scalar components with a mass much smaller than $M$,\footnote{A 
point similar to this last observation was already made in \cite{AHK}.} and the gravitino is also not light.

The content of the components equations of the superfield equation (\ref{eq1app}) is the same as the one displayed 
in eqs.~(\ref{phih0})--(\ref{Fh0}) for the rigid case. The first equation states again that the $\phi^h$ must adjust to 
values compatible with the assumption that all the $\psi_\alpha^h$ and $F^h$ vanish in first approximation, 
whereas the second and the third equations tell that $\psi_\alpha^h$ and $F^h$ must actually have small 
but non-vanishing vanishing values. As before, these suppressed components are important and cannot 
be neglected. The final result is then a supersymmetric effective theory that is accurate at leading order in 
$\partial^\mu/M$, $\psi^i/M^{3/2}$, $F^i/M^2$ and $U/M$, but again a priori not limited to small $\phi^i/M$. 

We have checked in a number of examples that the supersymmetric effective theory defined 
by the superfield equation $W_h = 0$ and the standard effective theory defined by the ordinary 
equations $V_h = 0$ do indeed approximately coincide under the assumptions mentioned above. 
For the scalar potential, in particular, the region in the space of scalar fields $\phi^l$ where the 
two theories match is now defined by three conditions:\footnote{The first two conditions are as 
before required to make sure that there is indeed a hierarchy between the light and heavy 
eigenvalues of the full supersymmetric mass matrix. The last additional condition is, as 
already explained, equivalent to the condition $U(\phi^l) \ll M $.}
\be
m(\phi^l),\mu(\phi^l) \ll M \,,\;\; F^l(\phi^l),F^h(\phi^l) \ll M^2 \,,\;\; m_{3/2}(\phi^l) \ll M \,,
\ee
In this case, it is not possible to perform analytic checks. The reason for this is that the validity of the 
approximation requires not only the $F^i(\phi^l)$ to be small, but also $U(\phi^l)$ (corresponding 
to $m_{3/2}(\phi^l))$ to be negligible as compared with the mass scale $M$. One has then one more condition 
than scalar fields, and this makes it impossible to re-express the deviation between the two effective 
potentials as a function of $F^l(\phi^l)$ and $U(\phi^l)$ instead of $\phi^l$. This reflects the fact that, 
as already mentioned, there generically exists a domain in field space where all the $F^l(\phi^l)$ are 
small, but in order to have in addition that also $U(\phi^l)$ is small in a non-empty portion of this domain, 
one needs in general to adjust some coefficients in the theory. Nevertheless, we performed 
a numerical point-by-point check for several non-trivial examples and verified that indeed our 
general conclusions hold true.

As in the case of global supersymmetry, one can redo the same analysis in the presence of 
light vector multiplets $V^a$. For the same reasons as before, one finds that even in this more 
general case one can use the simple equation $W_h=0$ to integrate out heavy chiral multiplets.

\section{Vector multiplets in global supersymmetry}\setcounter{equation}{0}

Let us consider now again the case of global supersymmetry, but including both chiral multiplets $\Phi^i$
and vector multiplets $V^a$, which are split into light ones $V^r$ and heavy ones $V^{x}$. For simplicity
we shall restrict to Abelian gauge fields, but the generalization to the non-Abelian case is straightforward.
Using the standard superspace formalism, requiring $n \le 2$ corresponds as before to neglect any dependence 
on supercovariant derivatives, with the only exception of those present by construction in the kinetic terms 
for the gauge fields. The theory can then be parametrized in terms of a real K\"ahler potential 
$K = K(\Phi^i, \bar \Phi^{\bar \imath},V^a)$, a holomorphic superpotential $W = W(\Phi^i)$ and a 
holomorphic gauge kinetic function $H_{ab}(\Phi^i)$. The Lagrangian takes the following form
\bea\label{lag3}
{\cal L} &=& \int \! d^4 \theta \, K + \int \! d^2 \theta \, W + \int \! d^2 \bar \theta \, \bar W \\
&\;& +\, \frac 1{64} \int \! d^2 \theta \, H_{ab}\, \bar D^2 D^\alpha V^a \bar D^2 D_\alpha V^b 
+ \frac 1{64} \int \! d^2 \bar \theta \, \bar H_{ab}\, D^2 \bar D_{\dot \alpha} V^a D^2 \bar D^{\dot \alpha} V^b \nn \,.
\eea
The local gauge symmetries associated with the heavy vector fields must be gauge fixed. The 
most convenient type of gauge fixing is the one in which some charged chiral superfield is fixed to 
some reference scale. In such a gauge the corresponding vector multiplet becomes then a general 
real vector multiplet, with all its components being physical. This way of proceeding allows to integrate 
out the heavy vector superfields at the superfield level. The exact superfield equations of motion for the 
heavy vector superfields $V^{x}$ are obtained by first rewriting the last two terms of the Lagrangian 
(\ref{lag3}) as $D$-terms by dropping two supercovariant derivatives, and then varying ${\cal L}$ with 
respect to $V^{x}$. This gives:
\bea\label{vec}
\a\a K_{x} + \frac 1{8} D^\alpha \Big(H_{xa} \bar D^2 D_\alpha V^a \Big) 
+ \frac 1{8} \bar D_{\dot \alpha} \Big(\bar H_{xa} D^2 \bar D^{\dot \alpha} V^a \Big)  = 0 \,.
\eea

The presence of a large supersymmetric mass for $V^x$ means in this case that around the 
value $V_0^x$ at which it is stabilized, the K\"ahler potential $K$ has a large second 
derivative $K_{x x'}(V_0^x)$ proportional to $M^2$. The first term in eq.~(\ref{vec}) 
dominates then over the others, and $V^x$ is approximately determined by the simple equation 
$K_x(V^x_0)=0$. The departure from this approximate solution is in this case found to be 
$\Delta V_0^x \sim {\cal O} (D^4V^a/M^2,\:D^4\Phi^m/M^2)$, where $\Phi^m$ denotes
all the chiral multiplets that have not been frozen by gauge fixing conditions. In our 
approximation, this correction can be neglected, since it would contribute only terms with 
$n\ge3$. For the terms coming from $H$, which already lead to terms with $n=2$ without 
extra supercovariant derivatives, this is obvious. On the other hand, for the terms coming 
from $K$, which can now lead to terms with $n=1$ due to the vector superfields, this is due
to the fact that the leading correction is proportional to $K_x$, which vanishes on the 
approximate solution. Summarizing, one can thus integrate out the superfields 
$V^x$ by using the simple vector superfield equation
\bea
K_x = 0 \,.
\label{eqvect}
\eea
This equation determines the heavy vector superfields as real functions of the 
light vector superfields and the chiral superfields plus their conjugates:
\bea
V^x = V_0^x(V^\alpha,\Phi^m,\bar \Phi^{\bar m}) \,.
\eea
The effective theory for the $V^\alpha$ and $\Phi^m$ is then obtained by plugging back 
this solution into the original Lagrangian. In the particular case in which
there are no light vector multiplets one needs to consider 
only $K$ and $W$, and one finds:
\bea
\a\a K^{\rm eff} (\Phi^l,\bar \Phi^{\bar l}) = K(\Phi^l,\bar \Phi^{\bar l}, V_0^x(\Phi^l,\bar \Phi^{\bar l})) \,,\nn \\
\a\a W^{\rm eff}(\Phi^l) = W(\Phi^l) \,.
\eea
In the case where there are also light vector multiplets, one can also get new effects from the gauge 
kinetic terms.

As in the case of chiral multiplets, the components of the superfield equation (\ref{eqvect}) have 
a simple interpretation. To spell it out, let us first notice that the supersymmetric mass matrix for 
vector superfields is given by $K_{ab}$, and the three blocks corresponding to the heavy ones,
the light ones and their mixing are thus given by:
\be
M^2_{xx'} = 2\,K_{xx'} \,,\;\; m^2_{rr'} = 2\, K_{rr'} \,,\;\; \mu^2_{xr} = 2\, K_{x r} \,.
\ee
The other object that enters is the couplings with chiral multiplets:
\be
q_{xi\bar \jmath} = K_{x i \bar \jmath} \,.
\ee
Notice next that eq.~(\ref{eqvect}) makes sense in any gauge. 
In order to interpret its components in physical terms, the most convenient choice is a 
supersymmetric gauge. In this way, one finds that the real scalar component $c^x$ 
must adjust its value in a way compatible with the approximate vanishing of $D^x$, 
whereas the other components are related to corresponding components of the light 
superfields through coefficients suppressed by inverse powers of the heavy mass. 
One may also go to the Wess-Zumino gauge to simplify the component expansion. 
The first few components of (\ref{eqvect}) imply then restrictions on the charged chiral 
multiplets fields. In particular, their scalar fields must adjust in such a way that the tadpole 
for $D^x$ cancels, whereas their auxiliary fields are subject to a linear relation corresponding
to the gauge invariance of the superpotential. The higher components of (\ref{eqvect}) 
imply on the other hand that the non-trivial components of the vector superfield can be 
reexpressed in terms of components of the charged chiral multiplets. In particular, 
one finds that 
\be
D^x = - (M_0^{-2}\! \mu_0^2)^x_{\;\,r} D^r - 2 (M_0^{-2}\! q_0)^x_{\;i \bar \jmath} \, F^i \bar F^{\bar \jmath} \,.
\label{Dx}
\ee
This equation coincides with the exact equation of motion of the complex
partner of the would-be Goldstone boson eaten by the gauge boson. Notice
that the first term has $n=1$ and is always relevant whereas the second 
gives $n=2$ and can thus give a relevant contribution only in $K$.

\section{Vector multiplets in local supersymmetry}\setcounter{equation}{0}

Let us finally consider the case of local supersymmetry, but including both chiral multiplets $\Phi^i$
and vector multiplets $V^a$ that are split into light ones $V^r$ and heavy ones $V^x$.
Using the superconformal superspace formalism, the requirement $n \le 2$ corresponds as before 
to simply neglect any dependence on supercovariant derivatives, except the ones in the kinetic 
terms for the gauge fields. 
The theory can then again be parametrized in terms of a real K\"ahler potential 
$K = K(\Phi^i, \bar \Phi^{\bar \imath},V^a)$, a holomorphic superpotential $W = W(\Phi^i)$ and 
a holomorphic gauge kinetic function $H_{ab}(\Phi^i)$ \cite{SUGRA2A,SUGRA2B}. 
The Lagrangian takes in this case the form
\bea\label{lv2}
{\cal L} &=& \int \! d^4 \theta \, \Big(\!-\! 3\, e^{-K/3} \Big) \bar \Phi \Phi 
+ \int \! d^2 \theta \, W \Phi^3 + \int \! d^2 \bar \theta \, \bar W \bar \Phi^3 \\
&\;& +\, \frac 1{64} \int \! d^2 \theta \, H_{ab}\, \bar D^2 D^\alpha V^a \bar D^2 D_\alpha V^b 
+ \frac 1{64} \int \! d^2 \bar \theta \, \bar H_{ab}\, D^2 \bar D_{\dot \alpha} V^a D^2 \bar D^{\dot \alpha} V^b \nn \,.
\eea
As in the rigid case, the local gauge symmetry associated to each heavy vector superfield
must be gauge-fixed, and the most convenient way to do this is to set a charged chiral superfield 
to some reference value. The exact superfield equations of motion for the heavy vector superfields 
are then obtained as before, and read:
\bea\label{kv2}
\a\a K_x + \frac 1{8} e^{K/3} (\bar \Phi \Phi)^{-1} \Big[D^\alpha \Big(H_{xa} \bar D^2 D_\alpha V^a \Big) 
+ \bar D_{\dot \alpha} \Big(\bar H_{xa} D^2 \bar D^{\dot \alpha} V^a \Big) \Big] = 0 \,.
\eea

The presence of a large supersymmetric mass implies again that around the values $V_0^x$ at 
which the heavy superfields $V^x$ are stabilized, the K\"ahler potential $K$ has a large second 
derivative $K_{x x'}(V_0^x)$ proportional to $M^2$. The first term in eq.~(\ref{kv2}) 
dominates then over the others, and $V^x$ is approximately determined by the equation 
$K_x(V^x_0)=0$. As before, the departure from this approximate solution is found to be 
$\Delta V_0^x \sim {\cal O} (D^4V^a/M^2,\;D^4\Phi^m/M^2)$, and can
be neglected. Summarizing, one can thus integrate out the superfields 
$V^x$ by using the same simple vector superfield equation as in the rigid case, namely
\bea
K_x = 0 \,.
\eea
Note that this equation is automatically and trivially invariant under K\"ahler transformations,
since these are not allowed to depend on the vector superfields. 

The components of this superfield equation admit exactly the same interpretation as in the global
case. This makes sense as long as the auxiliary field of the compensator is small, implying 
$m_{3/2} \ll M$. In particular, eq.~(\ref{Dx}) still holds true, but comparing it with the exact equation of 
motion of the complex partner of the would-be-Goldstone mode (see for instance \cite{CJ,GRSDT}
and also \cite{KK,K}),
one finds that it agrees with it only in the limit where $m_{3/2} \ll M$ and also $F^i \ll M$, which 
are indeed satisfied in our approximation.

\section{Conclusions} \setcounter{equation}{0}

In this paper, we have addressed the general question of understanding under which 
conditions it is possible to define a two-derivative supersymmetric low-energy effective 
theory by integrating out a heavy superfield with mass $M$, and with what procedure this theory 
can be explicitly computed. We studied the cases of chiral and vector multiplets, both in 
global and in local supersymmetry. Concerning the conditions for the existence of such 
a theory, we have argued that one has to require that all the derivatives, fermion fields and 
auxiliary fields should be small in units of $M$. In the global case, this means $\partial^\mu \ll M$ 
on all the fields, $\psi^i,\lambda^a \ll M^{3/2}$ for the chiralini and gaugini, 
and $F^i,D^a \ll M^2$ for the chiral and vector auxiliary fields. In the local case, one has in 
addition to impose $\partial^\mu \ll M$ on all the gravitational fields, $\psi_\alpha^\mu \ll M^{3/2}$ 
for the gravitino and $U \ll M$ for the gravitational scalar auxiliary field. This implies 
that $M$ should correspond to a supersymmetric mass, that comes from $W$ for chiral multiplets and 
from $K$ for the vector fields. We have then shown that under the above conditions the superfield 
equations allowing to integrate out heavy chiral and vector superfields $\Phi^h$ and $V^x$ 
in terms of light chiral and vector superfields $\Phi^l$ and $V^r$ are respectively 
the stationarity of the superpotential and the K\"ahler potential $W$ and $K$:
\bea
\begin{array}{l}
\partial_h W (\Phi^l,\Phi^h) = 0 \smallskip\ \\
\partial_x K (\Phi^l, \bar \Phi^{\bar l},\Phi^h,\bar \Phi^{\bar h},V^r,V^x) = 0 
\end{array}
\;\;\;\Rightarrow\;\;\; 
\begin{array}{l}
\Phi^h = \Phi_0^h (\Phi^l) \smallskip\ \\
V^x = V_0^x (\Phi^l, \bar \Phi^{\bar l}, V^r)
\end{array}
\;.\label{c12}
\eea
The fact that these equations are exactly the same in globally and locally supersymmetric
theories is a consequence of the assumption that higher-derivative terms should be negligible 
also in the gravitational sector. This implies that the gravitino 
mass should be much smaller than the supersymmetric mass $M$ of the superfield to be 
integrated out: $m_{3/2} \ll M$. One is then in a situation where the coupling to gravity is minimal and 
essentially dictated by the space-time symmetries, except for the Einstein term, which 
can however be canonically normalized in a universal way by going to the Einstein frame
from the start. As a result, the operations of integrating in/out heavy fields and 
switching on/off gravitational interactions commute. Exactly the same thing is true 
also for a generic non-supersymmetric theory, where the two-derivative effective 
theory can be deduced by integrating out the heavy fields by imposing stationarity of 
the potential.

In general, integrating out heavy superfields induces relevant corrections to the dynamics 
of the light superfields, which cannot be ignored in many interesting situations. In principle, 
to compute these corrections one simply needs to solve the superfield equations (\ref{c12}), 
which is a simple algebraic problem. In practice this may however be a non-trivial task, 
for example due to non-linearities or due to the proliferation of fields. It is then of 
interest to understand in which cases the effect of integrating out heavy superfields
is trivial, in the sense that it is equivalent to freezing these to some constant values
independent of the light superfields. According to the above superfield equations 
(\ref{c12}), we see that for chiral superfields this is the case when 
$W$ is separable, $W = W_L(\Phi^l) + W_H(\Phi^h)$, which still allows for 
non-trivial heavy-light interactions in $K$. On the other hand, for vector superfields 
one would need $K$ to be separable (for supersymmetric gauges), 
$K = K_L(\Phi^m, \bar \Phi^{\bar m}, V^r) + K_H(V^x)$, which implies that there 
are no heavy-light interactions at all since 
vector superfields are not allowed to appear in $W$. Actually, as far as the effective 
potential is concerned, this still approximately works even in the case where 
$W$ or $K$ respectively consist of a dominant term depending only on the heavy 
superfields and an other one depending also on the light ones but suppressed by 
some small parameter $\epsilon$, provided that the gravitino mass is at most of the 
same order $\epsilon$.\footnote{Note that this generically implies that $m_{3/2} \sim m$, 
which is a stronger condition than the restriction $m_{3/2} \ll M$ that is needed to be able 
to define a supersymmetric effective theory.} The reason is that $W$ and $K$ are stationary 
with respect to the approximate solution for the heavy chiral and vector fields and corrections 
can thus arise only at second order. This property was already derived in a different way
in \cite{GS}, and further generalizations will appear in \cite{GS2}. Notice however that there may 
be cases in which $W$ and/or $K$ are not separable but can be made separable after a 
superfield redefinition. In that case, the integration of heavy superfields will also be trivial, 
but only in the new superfield basis. Using the original field basis, one would find non-trivial 
corrections for the light field dynamics, but these clearly simply implement in an automatic 
way the field redefinition to the clever basis of light fields. On the other hand, in a generic 
effective theory a heavy field can be integrated out in a exactly trivial way only if the potential
$V$ is separable, at least at the point where the heavy fields are stabilized. For supersymmetric 
theories, to have such an exact trivialization one needs the stronger conditions that both 
$K$ and $W$ are separable in the rigid case, and that $K$ is separable and $W$ factorizable 
in the local case \cite{BDKV}, again at least at the point where the heavy fields are frozen 
\cite{GRS,AHS}.

As a final remark, let us emphasize that although the above results were 
derived at the classical level, similar considerations apply also at the quantum level.
In particular, it is always true that superfields with large supersymmetric masses can be 
integrated out at the level of superfields to define a two-derivative supersymmetric low-energy 
effective action. Due to the non-renormalization theorem for $W$, these loop corrections affect 
only $K$. See for example \cite{ISS} and \cite{RSS} for explicit examples in globally and locally 
supersymmetric theories.

\section*{Acknowledgments}

We would like to thank A.~Falkowski for collaboration during the initial stage 
of this project. We also thank R.~Rattazzi, A.~Romanino and M.~Serone for useful discussions.
This work was partly supported by the Swiss National Science Foundation.

\end{document}